\documentclass{optica-article}

\journal{opticajournal} 

\articletype{Research Article}
\usepackage{xcolor}
\usepackage{soul}
\usepackage{lineno}
\usepackage{float}

\begin{document}

\title{High-energy chirped nanosecond pulsed laser system for particle diagnostics and manipulation}

\author{Stefan Karatodorov,\authormark{1} Marios Kounalakis,\authormark{1}, Gabriel Flores Alfaro, \authormark{1,2}, Yingjie Zhao, \authormark{1,2}, Atulya Kumar, \authormark{1,2}, Ashwini Vaishnav, \authormark{1,2}, Jeremy Ramos, \authormark{1,2} and Alexandros Gerakis, \authormark{1,3,*}}

\address{\authormark{1}Luxembourg Institute of Science and Technology (LIST), 4422 Belvaux, Luxembourg\\
\authormark{2}Université du Luxembourg, 2 Av. de l’Université, L-4365 Esch-sur-Alzette, Luxembourg\\
\authormark{3}Department of Aerospace Engineering, Texas AM University, College Station, TX 77843,
United States of America}

\email{\authormark{*}alexandros.gerakis@list.lu} 


\begin{abstract*} 
An upgraded high-energy nanosecond pulsed laser system tailored for optical particle diagnostics and manipulation capable of pulse energies beyond Joule-level is presented.
In addition to the notable output energy increase, the laser system maintains its capability to generate laser pulses with customizable temporal profiles and variable durations ($1$~ns to $1$~$\mu$s) along with a chirping range of several GHz over the pulse duration.
The expanded output energy range is anticipated to greatly broaden the laser system's application potential, both for thermodynamic diagnostics via coherent Rayleigh-Brillouin scattering, by substantially lowering the particle density detection thresholds, as well as for particle manipulation, by facilitating more efficient optical trapping potentials for particle acceleration and deceleration. 

\end{abstract*}

\section{Introduction}
Frequency-agile lasers find use in a variety of fields, ranging from atomic, molecular and solid-state physics, to precision metrology and aerospace engineering~\cite{truong2013frequency,millot2016frequency,lihachev2022low,bak2022dual,derose2023high}.
This is due to their ability to generate pulses of stable intensity while having their output frequency varied as a function of time, i.e. being chirped, in a controlled, user-defined manner.
This feature can be used to expand the capabilities of optical methods based on laser-induced gratings~\cite{stampanoni2005gas,eichler2013laser}, which usually rely on the creation of a standing optical lattice, i.e a stationary periodic spatial density perturbation in a gas medium.
In particular, the incorporation of chirped intense pulses enables the creation of traveling optical lattices leading to advancements in optical probing/diagnostics, such as neutral gas and plasma diagnostics~\cite{pan2002coherent,pan2004coherent,gerakis2013single,bak2023torr,kumar2024multi,kumar2025multi}, as well as optical manipulation of matter, e.g. trapping, probing and acceleration/deceleration of atoms and molecules~\cite{wieman1999atom,ertmer1985laser,greiner2002quantum,Savall1999optical,greiner2008optical,jones2002moving,cadoret2009atom,dong2005cold} and antihydrogen transport~\cite{barker2012directed}. 

However, many of these applications impose non-trivial requirements on the output parameters of the laser sources needed for their implementation, i.e. rapid chirp rate (> 1~MHz/ns), high pulse energy, high temporal intensity stability, variable pulse durations (ns-$\mu$s),  variable temporal pulse profile, multiple output wavelengths; these requirements alone or combined often exceed the capabilities of commercially available lasers.
Therefore, several efforts have focused on developing laser systems that meet these criteria, e.g. the laser systems developed in Refs.~\cite{coppendale2011high} and~\cite{bak2022dual}, which were subsequently used for the acceleration of Ar metastable atoms to a maximum velocity of $191$~m/s~\cite{maher2012laser} and for thermodynamic single-shot diagnostics of gas particles (atomic and molecular) at single-Torr level pressures~\cite{bak2023torr}, respectively. While these efforts have addressed some of the necessary requirements, the development of chirped laser sources  of higher pulse energy (exceeding the level of $1$~Joule/pulse) is of crucial importance, e.g. for probing lower gas densities (order of $10^{-3}-10^{-4}$ Torr in air) for space applications and low temperature plasma diagnostics or for optical manipulation with deeper optical trapping potentials facilitating particle acceleration to higher maximal velocities (beyond hundreds of ms$^{-1}$).

In the present work, a chirped laser system capable of Joule-level pulse energy output meeting the previously outlined requirements is presented for the first time to the best of the authors' knowledge. This laser system exhibits a five-fold increase of the output pulse energy compared to the previous iteration~\cite{bak2022dual}.
An additional two-pass amplification stage is designed, built and commissioned, which is discussed in detail along with the encountered technical challenges and implemented solutions.
In addition to the energy increase, it is demonstrated that the laser system preserves its unique capabilities in terms of fast chirp rates of up to $\sim\pm2.5$~GHz, variable pulse duration ($1$~ns to $1$~$\mu$s), arbitrary pulse temporal profile and Gaussian output beams.
Finally, the increased capabilities of the resulting high-energy laser source are showcased by its application for neutral gases 
using coherent Rayleigh-Brillouin scattering (CRBS).  

\section{Experimental setup: chirped laser system upgrade to Joule level pulse energy}

In this section, the design and commissioning of the chirped high-energy pulsed laser system is presented. This work builds upon an existing laser source, described in detail in Ref.~\cite{bak2022dual}, which is briefly outlined in Section~\ref{sec:4th-stage}.
The upgrade of the laser system to Joule-level pulse energies, incorporating an additional amplification stage, is discussed thoroughly in Section~\ref{sec:5th-stage}. The output capabilities of the upgraded laser source are  extensively characterized in Section~\ref{sec:output-characterization}.

\subsection{\label{sec:4th-stage}Frequency, pulse duration and shape agile laser system for particle spectroscopy and manipulation with dual-color output}

\begin{figure}
\centering
\includegraphics[scale=0.40]{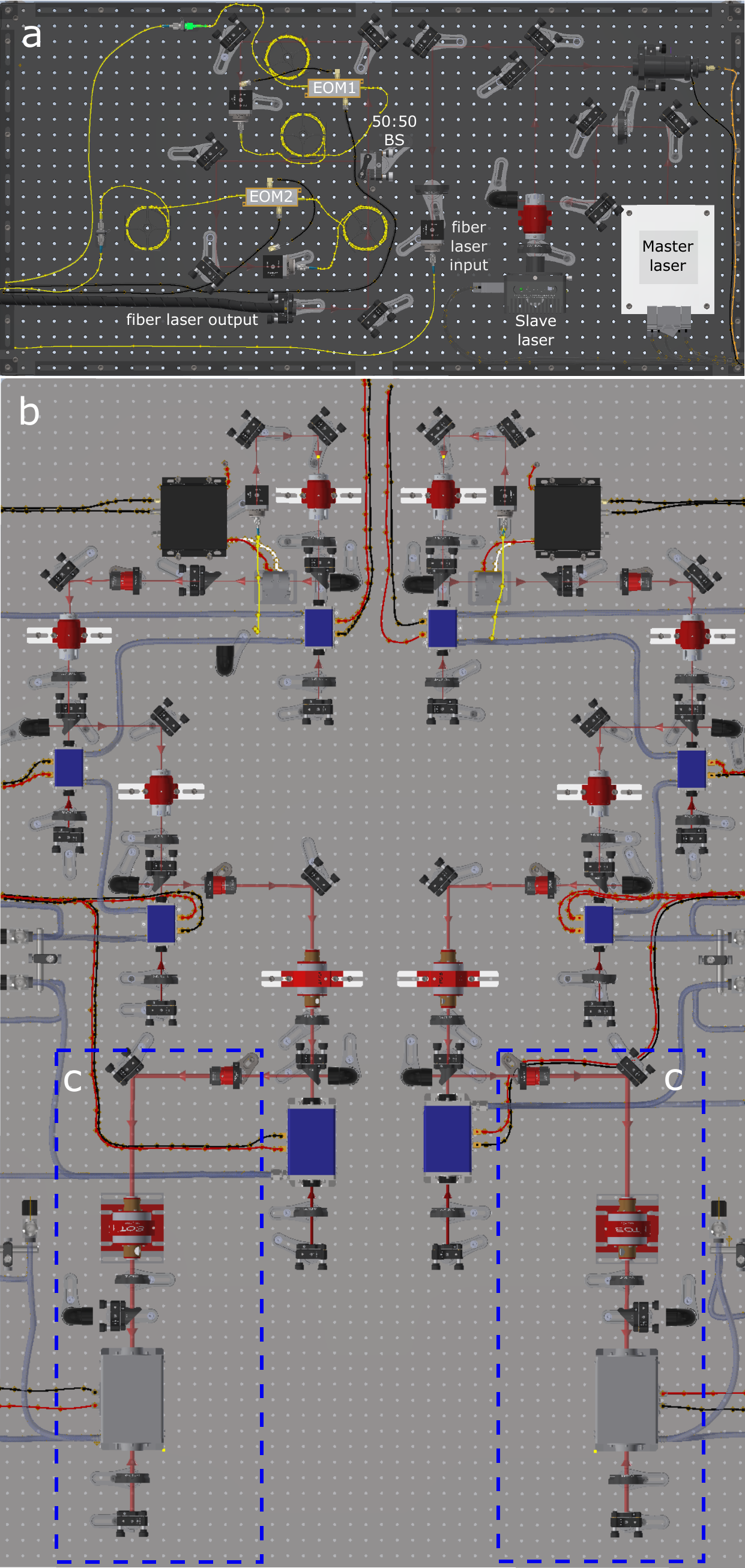}
\caption{\label{fig:optical-table} Top-view digital twin of the laser system presented here: (a) master-slaver laser and the fiber amplifier breadboard, and (b) all the amplification stages of the laser system (blue units) with (c) the added amplification stages (indicated by dashed blue rectangles).}
\end{figure}

The laser setup~\cite{bak2022dual} begins with a custom-built continuous wave (CW) Nd:YVO$_4$ microchip laser emitting at $1064$~nm, that is capable of rapid output frequency modulation (chirp) through the use of an intracavity LiTaO$_3$ crystal.
The purpose of this \textit{master} laser is to generate pairs of un-chirped and chirped CW laser output that is subsequently shaped in two separate pulses of equal duration having a relative time delay, $\Delta t_\mathrm{delay}$, between them.
The output of the master laser is injection-locked onto a $1064$~nm emitting diode laser\footnote{M9-A64-0200, Thorlabs}.
The function of this \textit{slave} laser is to bypass the relaxation oscillations occurring at the master laser when its output frequency is rapidly changed~\cite{siegman1986lasers,koechner2013solid}.
The slave laser output is coupled to a fiber amplifier\footnote{YAR-10-1064-LP-SF, IPG Photonics} for amplification to $1$~W, and subsequently split into two beams by a $50$:$50$ beamsplitter, as depicted in Fig.~\ref{fig:optical-table}(a).
Each beam passes through an intensity electro-optical modulator (EOM)\footnote{AM1064b, Jenoptik}, controlled by an arbitrary waveform generator (AWG)\footnote{AFG3102C, Textronix}; this enables the individual shaping of the un-chirped and chirped parts of the CW beam into two separate laser pulses with preset duration ($1~\mathrm{ns}-1~\mu$s) and arbitrary temporal profile.
The pulses are then transmitted through two optical fibers with different lengths to compensate for the delay, $\Delta t_\mathrm{delay}$, between them so that they are temporally aligned at the beginning of the pulse amplification chain (Fig.~\ref{fig:optical-table}(b)).
Each of the two pulse beams is subject to a respective four-stage amplification chain composed of Nd:YAG diode-pumped amplifiers \footnote{RBAT20, two RBAT34 and REA7006, Northrop Grumman Cutting Edge Optronics} that increases the initial energy from $100$~nJ up to $450$~mJ per pulse, with a beam diameter of $7.5$~mm.

\subsection{\label{sec:5th-stage}High-energy pulsed laser system upgraded with additional amplification stage}
The main focus of this work is to further enhance the laser system capabilities by increasing its output energies beyond $1$~J per pulse per arm, which is enabled by the incorporation of an additional two-pass amplification stage on each laser arm.

\subsubsection{\label{sec:design} Design and operation}
The amplification stage design includes a variable beam expander, a $\lambda/2$ waveplate, a Faraday isolator, a thin-film polarizer (TFP), a diode-pumped amplification module~\footnote{REA15008, Cutting Edge Optronics} cooled by a dedicated chiller~\footnote{CA05A2T3-41CAIN, PolyScience},  a $\lambda/4$ waveplate, and a $0^{\circ}$ angle of incidence (AOI) mirror, as illustrated in Fig.~\ref{fig:optical-table}(c).

The beam expander is utilized in order to magnify the laser beam from the fourth amplification stage to the maximum diameter permitted by the clear aperture of the isolator and the amplifier ($\sim80$\% of the amplifier rod diameter as recommended by the manufacturer).
The Faraday isolator ensures uni-directional transmission of the laser beam, thereby preventing feedback of the amplifier laser emission back to the previous stages of the amplification chain.
In addition, a $\lambda/2$ waveplate in series with a TFP are placed before the entrance to the amplifier, while a $0^{\circ}$ AOI mirror is placed at the exit.
This configuration enables a double-pass of the beam through the amplification stage. 
More specifically, following a first pass through the amplifier, the polarization of the beam is rotated by $90^{\circ}$ after passing twice through the $\lambda/4$ waveplate, resulting in the amplified beam being reflected out of the amplification stage by the TFP. 

\subsubsection{\label{sec:design} Installation and testing}
\begin{figure}[H]
\centering
\includegraphics[scale=0.23]{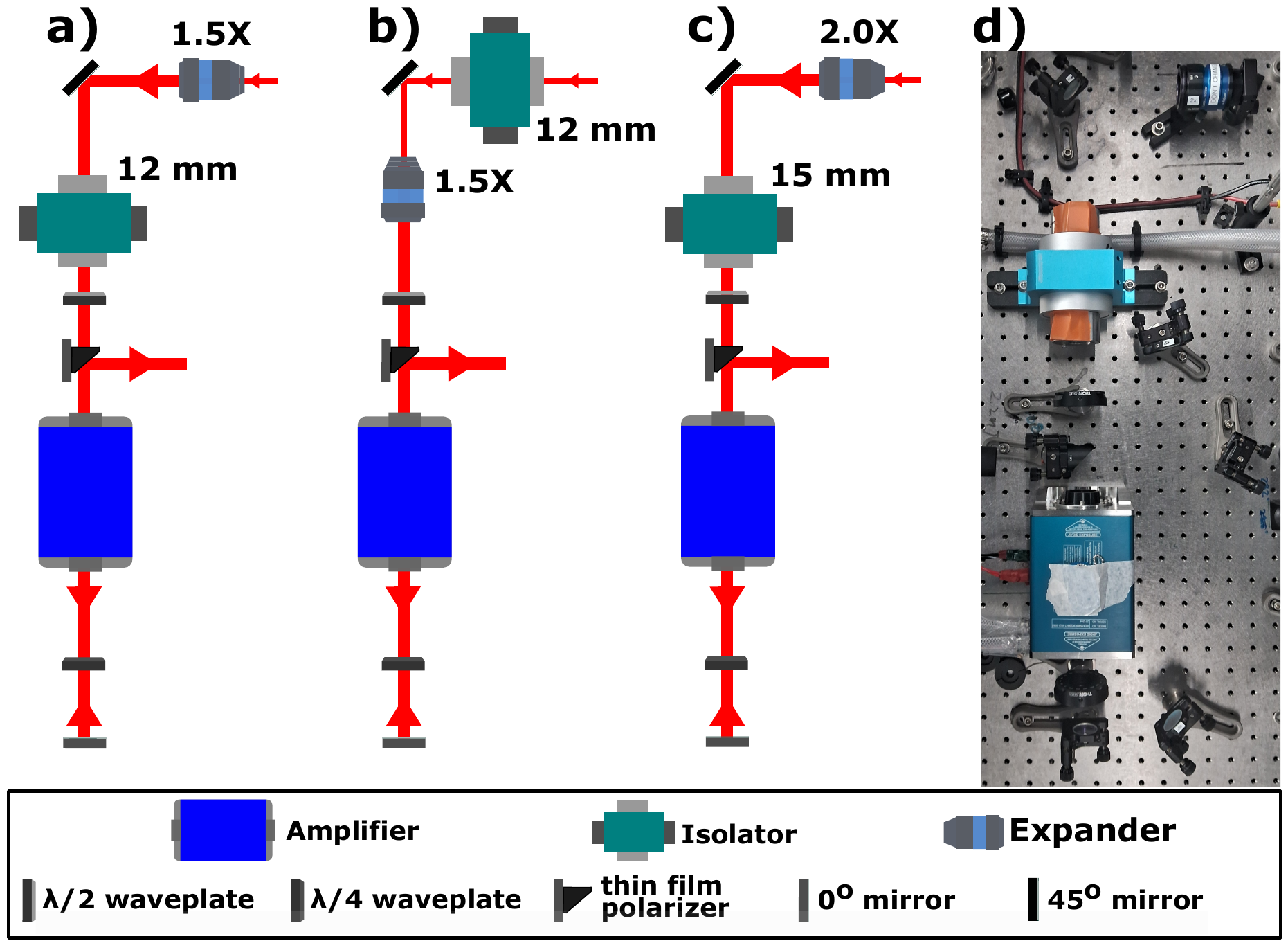}
\caption{\label{fig:conf} Three amplification stage configurations with: (a) a 12 mm clear aperture isolator and a 1.5$\times$ expander, (b) swapped positions of the isolator and 2$\times$ expander and (c) a 15 mm clear aperture isolator and 2$\times$ expander. (d) Photographic image of the implemented amplification stage depicted in (c).}
\end{figure}

The amplification stage design and implementation underwent multiple iterations prior to reaching optimal operational configuration, which is showcased in Fig.~\ref{fig:conf}(c) .
\newline
In this configuration, a $7$~mm diameter beam is incident on a 2$\times$ magnification~\footnote{161-2X-1H, Eksma Optics} expander which resulted in a collimated $14$~mm diameter beam. This beam is then directed into a $15$~mm clear aperture Faraday isolator\footnote{HPISO-FS-50-15-1064-N-A61, Castech}. The choice of this isolator model was motivated by the large clear aperture and higher laser-induced damage threshold of $10$~J/cm$^2$. With this configuration, a maximum output energy of $2.4$~J is obtained from an input energy of $0.41$~ J, at a $10$~Hz repetition rate, i.e. $5.36\times$ amplification. To the best of the authors' knowledge, this is the highest reported energy per pulse generated by a frequency-chirped laser source tailored for laser-induced dynamic optical grating applications at nanosecond timescales. For more information on the initial iterations shown in Fig.~\ref{fig:conf}, the reader is directed to Appendix~\ref{append::Amp design variations}.

\subsubsection{\label{sec:beam-spot}Amplification stage optimization}

\begin{figure}
\centering
\includegraphics[scale=0.12]{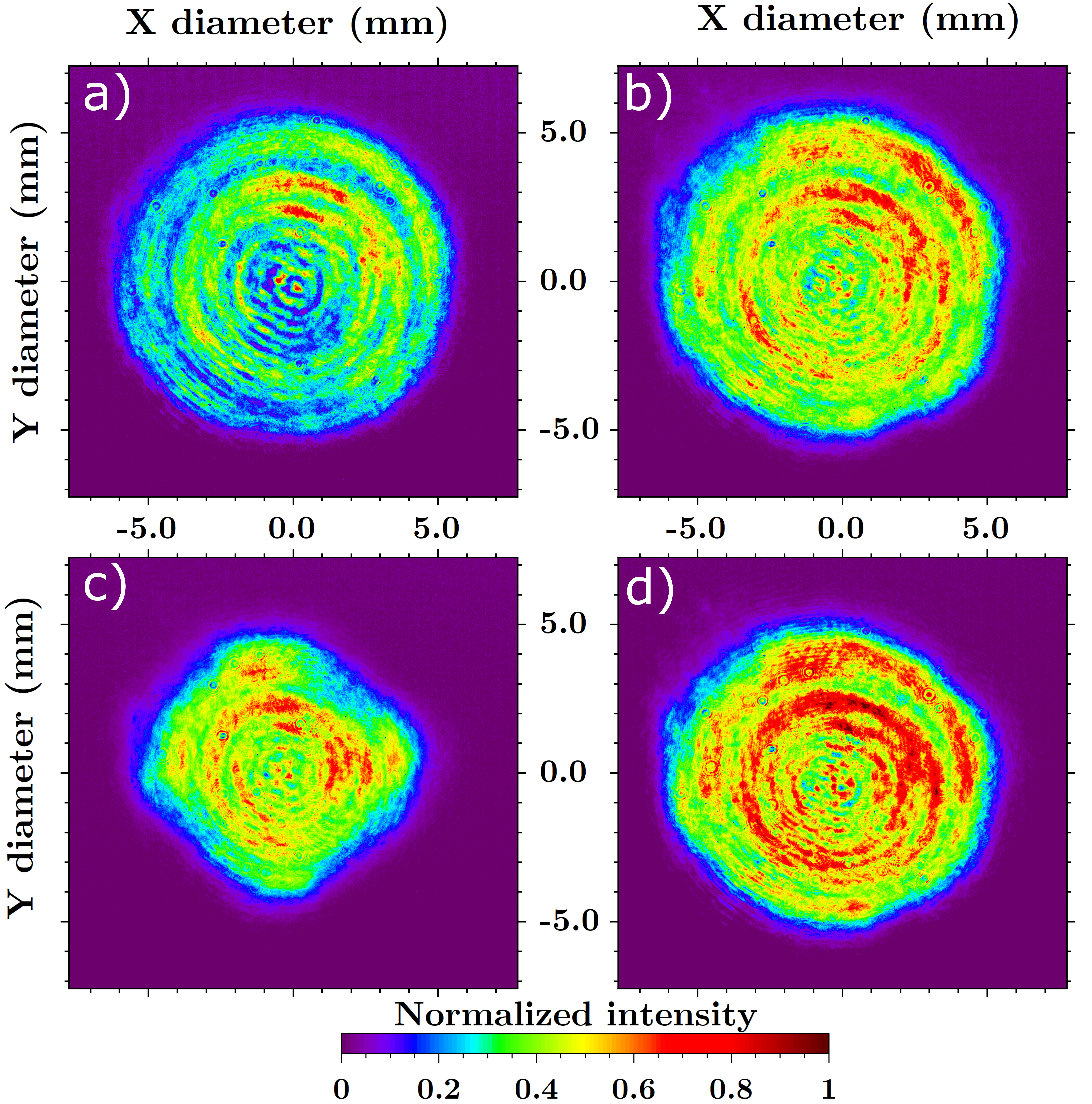}
\caption{\label{fig:beam-spot} Beam spot intensity distributions of (a) $14$ mm beam coming from the fourth amplification stage after passing through the 2$\times$ expander, (b) $14$ mm amplified beam at 10 Hz repetition rate during the first few seconds of operation, (c) distorted amplified beam at 10 Hz repetition rate after tens of seconds of operation, and (d) $14$ mm amplified beam at 5 Hz repetition rate.}
\end{figure}

Operating the laser system at such high pulse energies leads to unforeseen complications, which need to be addressed for the optimal operation of the laser system.
More specifically, a significant distortion in the beam spot intensity distribution together with a $\sim25\%$ reduction in the output pulse energy was observed after a few tens of pulses, when operating at a $10$ Hz repetition rate.
The distorted beam spot, shown in Fig.~\ref{fig:beam-spot}(c), is the result of thermally-induced birefringence from thermal stress accumulation in the laser crystal rod due to inadequate, non-uniform cooling~\cite{koechner2013solid}. This results in thermal lensing and subsequent depolarization of the laser beam.

An effective solution to mitigate this issue is to balance the thermal load on the crystal with the cooling rate.
However, due to technical limitations of the chiller used to cool the laser, it is not possible to increase the coolant flow rate.
Therefore, the adopted solution is to reduce the thermal load on the crystal by lowering the operational repetition rate of the laser, which results in a more efficient cooling rate of the crystal and a more stable laser operation.

\begin{figure}
\centering
\includegraphics[scale=0.35]{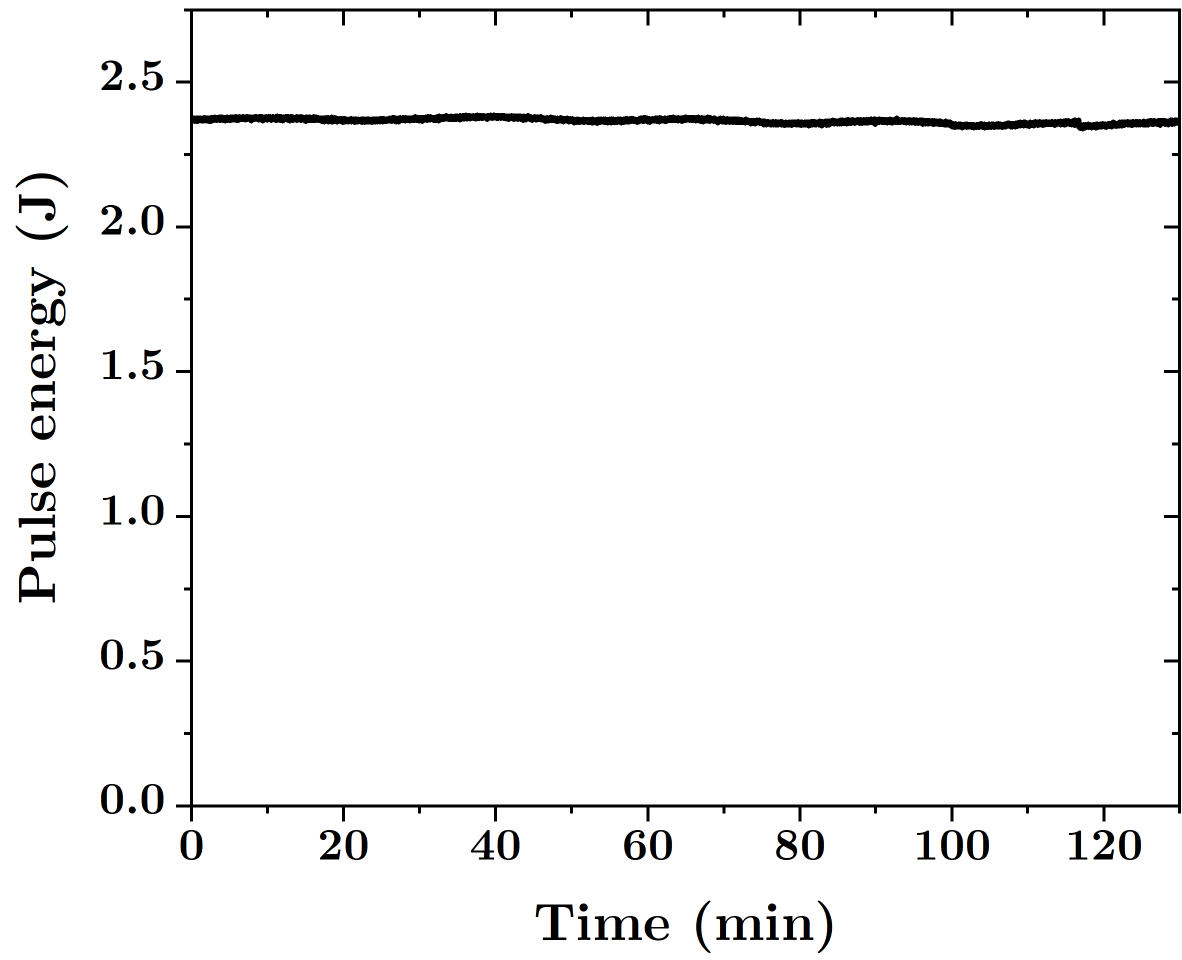}
\caption{\label{fig:energystability}Laser pulse energy monitored over a period of $\sim2$ hours.}
\end{figure}

It was experimentally confirmed that operating at a $5$~Hz repetition rate, led to no measurable change in the beam spot profile, as shown in Fig.~\ref{fig:beam-spot}(d), even for  prolonged operation of the laser.
Additionally, the laser energy ($2.36$~J on average) remains stable within $0.4\%$ at this lower repetition rate over several hours, as seen in Fig.~\ref{fig:energystability}.

\section{\label{sec:output-characterization}High-energy laser system characterization: pulse temporal profile, pulse chirp and second harmonic generation}

One of the most important requirements for the operation of the laser system is that the amplified laser pulses preserve the properties at the pre-amplification stage, i.e. the chirp applied by the master laser, and the temporal duration and profile of the pulses, shaped by the AWG and EOMs.
In order to demonstrate this, the properties of the amplified laser pulses have been characterized.

\subsection{\label{sec:chirp}Laser pulse chirp}

\begin{figure}[H]
\centering
\includegraphics[scale=0.6]{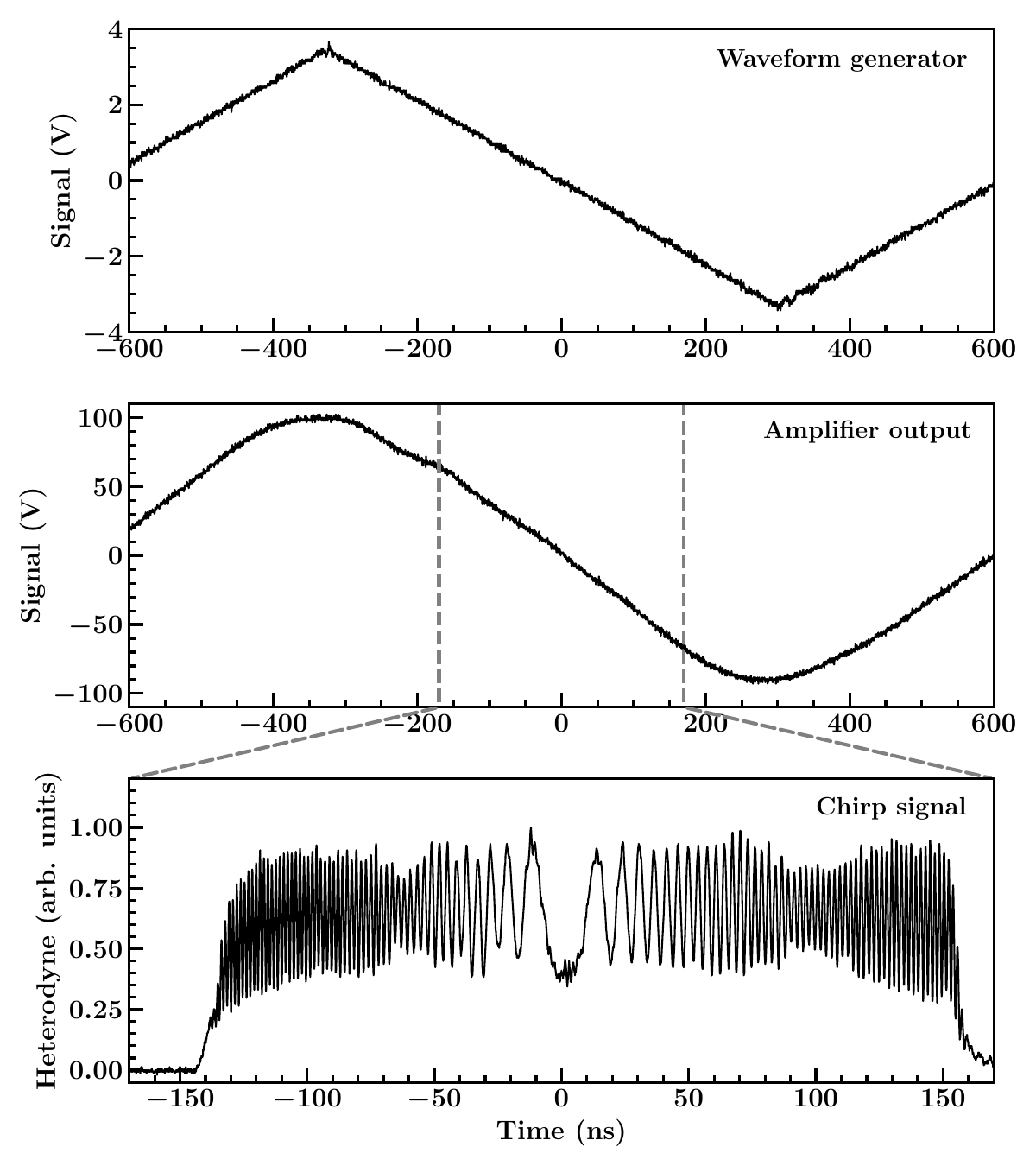}
\caption{\label{fig:chirp_distort}Chirp signal amplifier performance: (top) waveform generator input signal, (middle) amplified output and (bottom) generated laser pulse chirp on the heterodyne photo-diode. Note that the time scale in the bottom sub-figure is different, in order to show the laser pulse chirp in more detail, and the gray dashed lines in the middle sub-figure outline the boundaries of the magnified time window.}
\end{figure}

As discussed in Section~\ref{sec:4th-stage}, an important characteristic of this laser source is the generation of a pair of pulses having a relative frequency difference between them over the pulse duration, i.e. a chirp.
The profile of the chirp as a function of time, $\Delta f(t)$, can exhibit any structure set by the user. To demonstrate this, a V-shaped chirp structure in the frequency domain is shown (Fig.~\ref{fig:chirp_distort}), starting with a high frequency difference, decreasing to zero, and then increasing back to the same high frequency; this chirping operation occurs within the duration of the produced pulses, typically $150-300$~ns.
For diagnostic applications, the two pulses typically interfere within a medium to produce an optical lattice. The velocity of this optical lattice is given by $v_{lattice} = (\Delta f \lambda_{laser})/(2\sin(\phi/2))$, where $\lambda_{laser}$ is the laser wavelength and $\phi$ is the angle between the two interfering beams.
Therefore, the applied chirp leads to the lattice decelerating from a high velocity to zero and then accelerating back to the high velocity within the duration of a single laser pulse.
This property is crucial for probing the complete velocity distribution function of the gas or nanoparticle environment in single-shot CRBS, as demonstrated in Refs.~\cite{gerakis2013single,gerakis2016remote, gerakis2018four, kumar2024multi, kumar2025multi}.

\begin{figure}[H]
\centering
\includegraphics[scale=0.3]{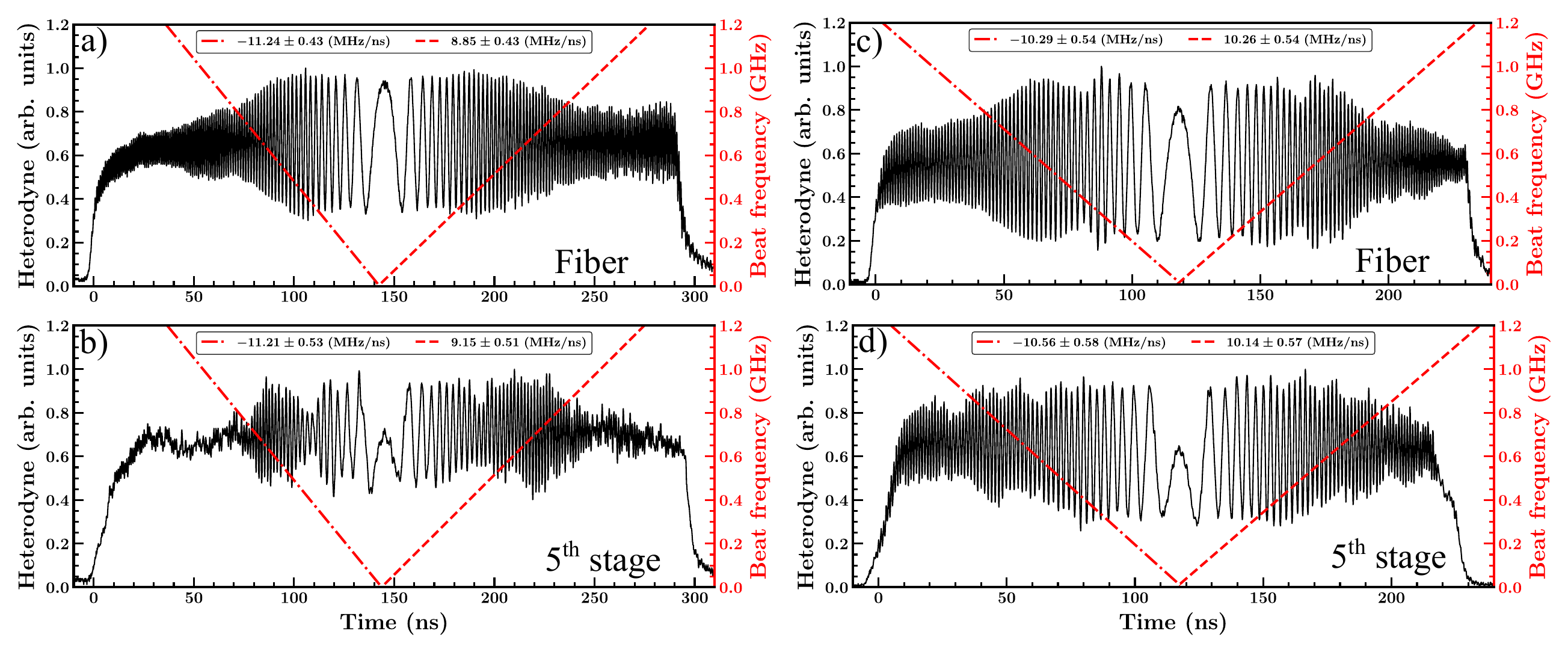}
\caption{\label{fig:chirp_fin} Heterodyne time-domain measurements (solid black curves) overlayed with the resulting frequency chirps (dashed and dashed-dot red lines), obtained after the fiber laser and the 5$^{\text{th}}$ amplification stage for two different master-slave injection locking positions: (a)-(b) sub-optimal and (c)-(d) optimal injection-locking.}
\end{figure}

The characterization of the relative frequency difference between the two pulses occurs with their beating and subsequent heterodyne detection on a fast photo-diode~\footnote{G6854-01, Hamamatsu Photonics}.
The recorded data for the pulse beating and thus the chirp are shown, before and after amplification by all five stages in Fig.~\ref{fig:chirp_fin}.
The solid black curves represent the heterodyned signal, while the dashed/dashed-dot red lines show the instantaneous frequency. The chirp rate is derived using the method described in Ref.~\cite{fee1992optical}, which agrees with the applied V-shaped frequency modulation profile.
When comparing the chirp rates before pulse amplification (Figs.~\ref{fig:chirp_fin}(a),(c)) and after pulse amplification (Figs.~\ref{fig:chirp_fin}(b),(d).), their difference is found to be minimal ($< 0.3$ MHz/ns) for the two locking states considered.
Consequently, the amplification process preserves the structure and magnitude of the applied frequency modulation.

Importantly, even if the chirp rate is preserved at different master/slave injection-locking positions, the beat pattern after pulse amplification could be drastically different (see Figs.~\ref{fig:chirp_fin}(b),(d)) due to the variable gain profile, i.e. not flat-top across the gain bandwidth, of the amplification crystals. By changing the master/slave locking position, the output laser frequency is changed; when the resulting beam is going through the amplification crystals the beam's new frequency (and its excursion across the pulse duration) will fall at different spectral positions within the arbitrarily shaped gain profile, thus experiencing frequency-dependent amplification~\cite{martial2011NdYAG, koechner2013solid}. Therefore, precise tuning of the output frequency is necessary, so that the resulting amplification is as close to a flat-top behavior as possible (e.g. compare Figs.~\ref{fig:chirp_fin}(b) and (d)). A locking position that preserves the beating contrast after amplification is thus referred to as optimal injection-locking, as in Figs.~\ref{fig:chirp_fin}(c)-(d), as opposed to a sub-optimal injection-locking, shown in Figs.~\ref{fig:chirp_fin}(a)-(b).

 Apart from the injection-locking, another source of frequency instability can be found through the voltage signal itself. The voltage signal that is applied to the intracavity LiTaO$_3$ crystal, originates from the output of a function generator~\footnote{3314A Function Generator, HewlettPackard} that is subsequently amplified by a high-voltage amplifier~\footnote{A-303 High Voltage Amplifier, A.A. Lab systems}. Fig.~\ref{fig:chirp_distort} shows a comparison between the input signal for the amplifier and its output signal along with the generated chirp. As seen in the figure, the amplifier distorts the saw-tooth input profile which in this particular case is more pronounced in the left part of the oscillation. This distortion can be minimized by applying an offset to the amplifier output. In order to rectify this effect, a customized high-voltage amplifier that will provide with an undistorted sawtooth amplified waveform is currently under development.

\subsection{\label{sec:temporal-profile}Laser pulse temporal profile}

\begin{figure}[H]
\centering
\includegraphics[scale=0.55]{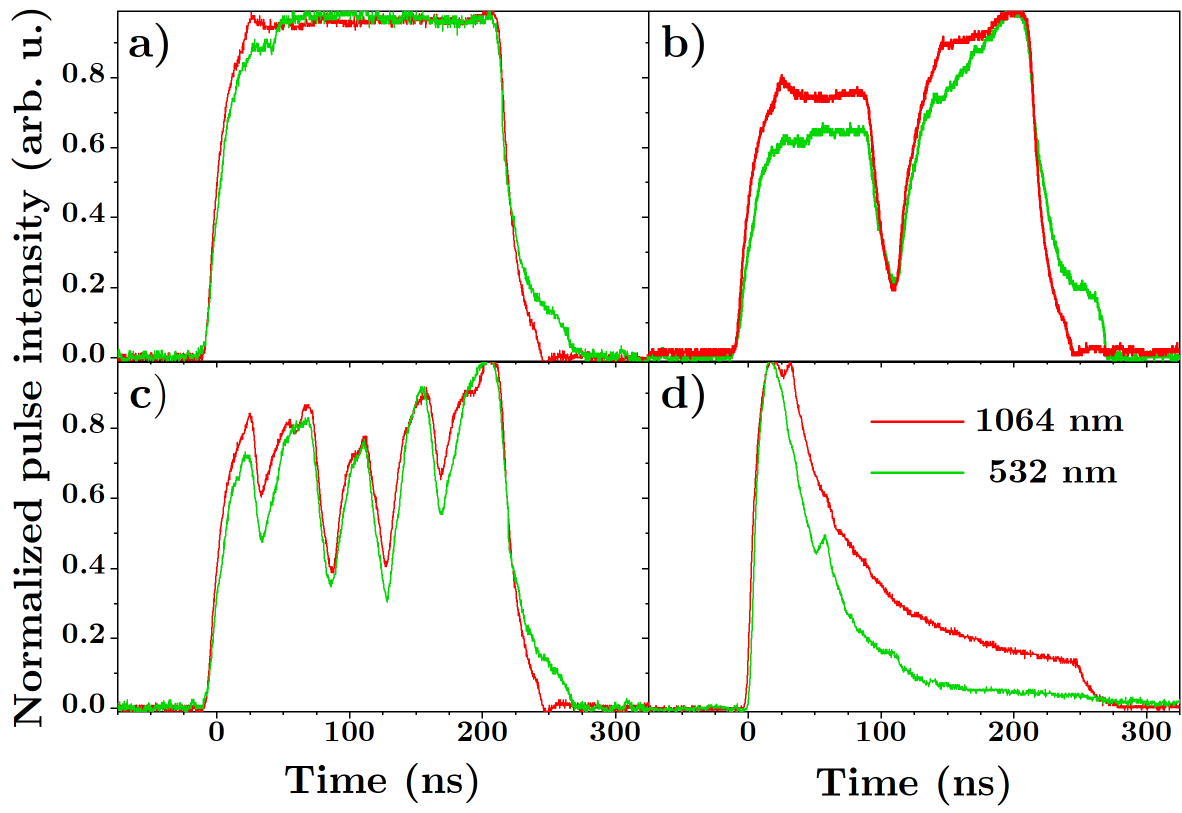} 
\caption{\label{fig:pulse_profile} Arbitrary temporal  pulse profiles generated after amplification at 1064 nm and 532 nm.}
\end{figure}

The previous version of this laser system, described in Section~\ref{sec:4th-stage} and Ref.~\cite{bak2022dual}, already permits control over the laser pulse temporal profile, enabling the creation of arbitrary pulse shapes.
This is achieved by adjusting the EOM transmission with a control signal from the AWG. The AWG is driven by a custom-made LabVIEW program that can modify the signal waveform to achieve the desired pulse shape. In this work, the LabVIEW program's capabilities have been expanded to automatically generate a flat-top pulse, which is the required profile for single-shot CRBS~\cite{gerakis2013single}.
It is demonstrated here that the high-energy upgrade of the laser system has maintained the ability to produce arbitrary temporal profiles.
Fig.~\ref{fig:pulse_profile} shows several arbitrary pulse profiles, including a flat-top one, generated with the amplified pulses.

\subsection{\label{shg}Second harmonic generation}

\begin{figure}[H]
\centering
\includegraphics[scale=0.28]{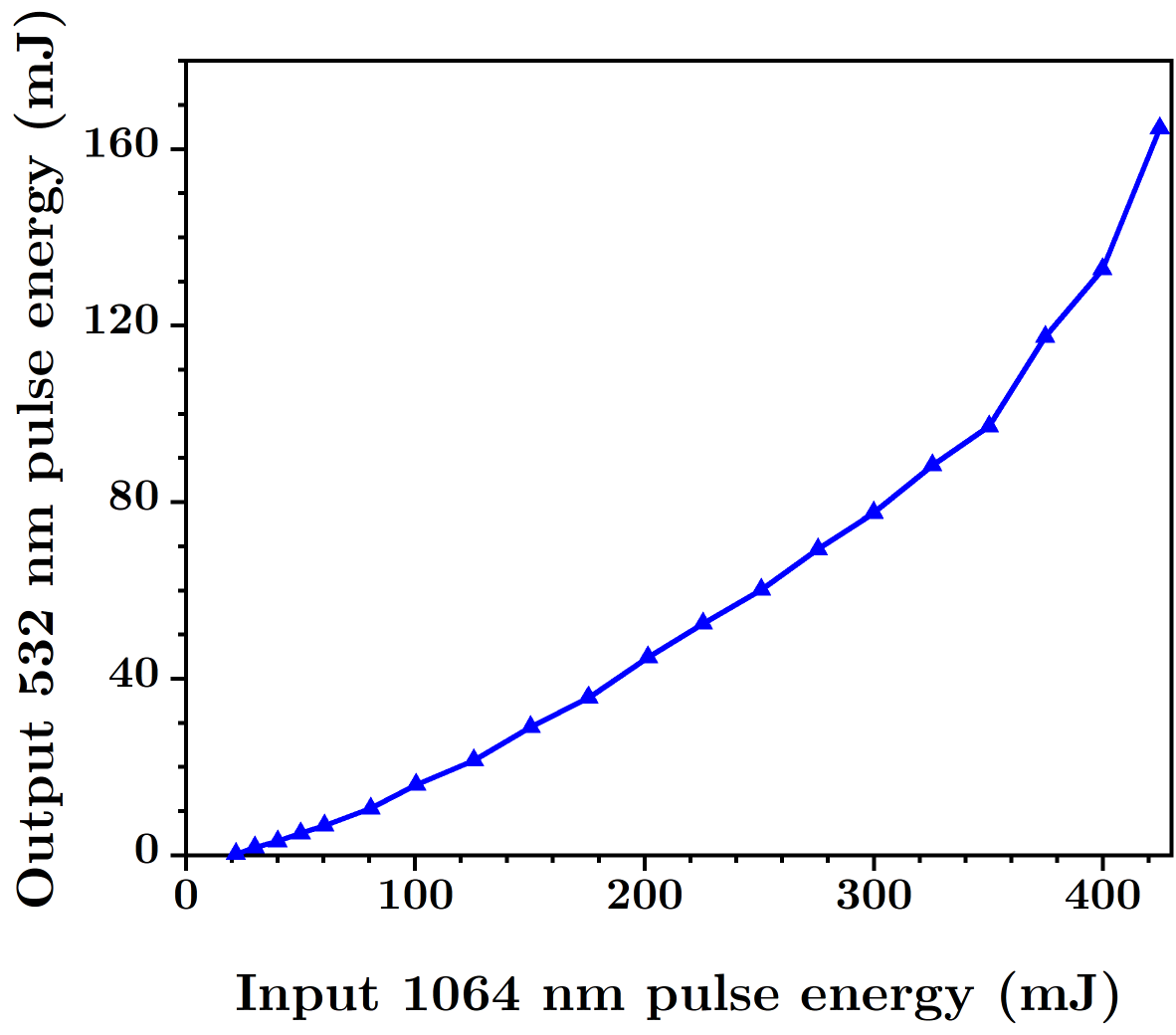}
\caption{\label{fig:shg}Second harmonic generation using the laser beam after the 5th stage amplification. Every point on the curve is an average of 50 laser pulses.}
\end{figure}

Part of one of the laser beams can also be used for the generation of pulses at the second harmonic, i.e. at a $532$~nm wavelength, which is particularly useful for applications where the dual color CRBS geometry is implemented~\cite{bak2023torr}.
This is due to the fact that the Rayleigh scattering cross-section scales as $\lambda_{probe}^{-4}$, thereby resulting in a $16$ times increase in the scattered signal. Additionally, more efficient detectors can be used at $532$~nm in comparison to the ones for $1064$~nm.
Both of these advantages enable the use of single-shot CRBS at lower gas densities (order of $10^{-1}$ Torr).
For this purpose, the second harmonic generation (SHG) setup discussed in Ref.~\cite{bak2022dual} is employed here.
The efficiency of the SHG has been measured for input pulse energies in the range of $10$~mJ to $450$~mJ and the results are shown in Fig.~\ref{fig:shg}. Every point on the curve of the second harmonic pulse energy represents an average of 50 individual laser pulses.
As seen in the figure the maximum achieved conversion efficiency is $38.7$\%.

\subsection{\label{sec:output-characterization}High-energy chirped laser system applications}
\begin{figure}[H]
    \centering
    \includegraphics[width=0.5\textwidth]{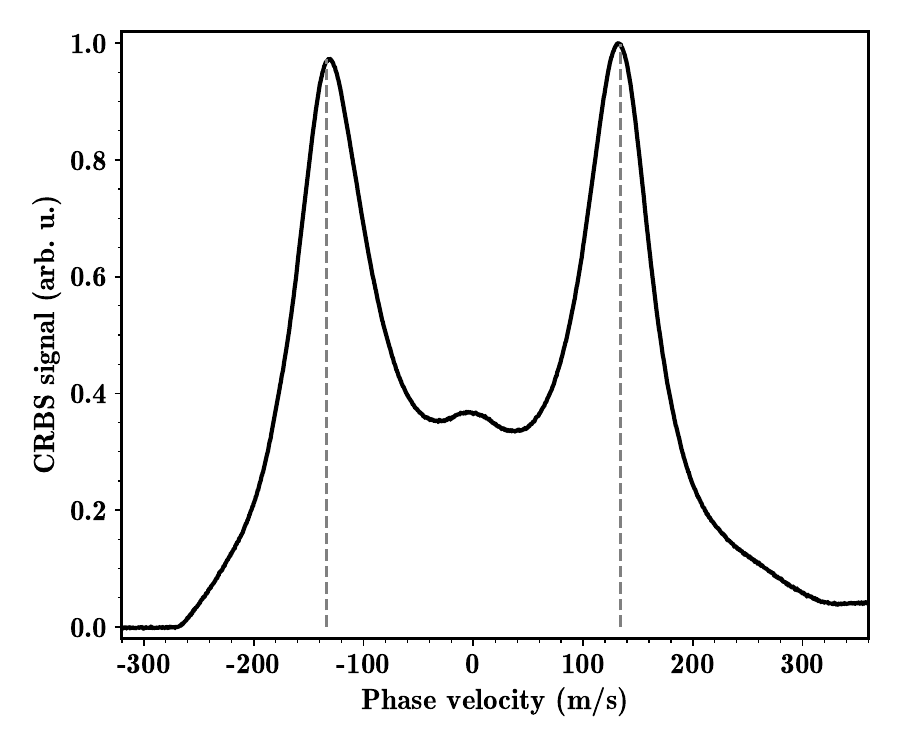}
    \caption{\label{fig:CRBS} CRBS spectrum for SF$_6$ gas at room temperature and a pressure of $1$ atm, obtained after $10$ averages.
   The Rayleigh peak is centered at zero phase velocity, while the Brillouin peaks are centered around $132$~ms$^{-1}$, which is within $2\%$ of the speed of sound in SF$_6$ measured at NIST ($134$~ms$^{-1}$, indicated by the dashed lines) at the same temperature and pressure.}
\end{figure}

The main objective for the commissioning of the laser pulse energy increase of the chirped laser system is its application for advanced laser diagnostics of neutral and charged gas environments and optical matter manipulation.
To demonstrate the capabilities of the system, it is employed for CRBS analysis of molecular gas (SF$_6$) at standard conditions. A CRBS spectrum of the analysis is shown in Fig.~\ref{fig:CRBS}. It is seen in the figure, that the Brillouin peaks are centered at 132~ms$^{-1}$, which is within $2\%$ of the speed of sound in SF$_6$ given at NIST (134~ms$^{-1}$).
In addition, the Rayleigh peak is visible, centered at zero phase velocity. 

\section{Conclusion}
In summary, a nanosecond chirped pulsed laser system capable of producing Joule-level energies per pulse has been demonstrated.
The laser system enables the production of high-energy chirped pulses at a fast rate of up to $27$~MHz/V and variable pulse durations ranging from a few nanoseconds up to microseconds.
A maximum energy output of $\sim 2.4$~J per pulse is demonstrated, while maintaining the main spatio-temporal characteristics of the pulsed beams. The repetition frequency of the laser is limited to $5$~Hz to provide efficient cooling of the fifth amplification stage amplifier rods resulting in distortion free beams spot distribution. At this repetition frequency, the laser system emission is monitored over several hours showing stability within $0.4\%$.

The unique capabilities of this system make it a valuable tool for optical laser particle diagnostics and manipulation, especially for pulsed operations relying on nanosecond-scale frequency tuning, such as the single-shot CRBS technique.
\section{Appendix}
\subsection{\label{append::Amp design variations}Amplification design variations}
The amplification stage design and implementation underwent three iterations prior to reaching the optimal configuration, as showcased in Figure~\ref{fig:conf}.
The differences between all three implementations lie in varying the relative positions of the expander and the isolator in addition to employing different optical parts, as explained below.

The first iteration, shown in Fig.~\ref{fig:conf}(a), employed a 1.5$\times$ expander~\footnote{EKSMA 161-1.5X-1H} placed before a $12$~mm clean aperture isolator~\footnote{Pavos EOT 12-01428}. The initial beam diameter was magnified to $11$~mm to fit the clear aperture of the isolator and to avoid any significant diffraction effects. In this configuration, laser output energies on the order of $1.3$~J per pulse were achieved from an initial input of $0.35$~J, corresponding to a 3.8$\times$ amplification at a $10$~Hz repetition rate. Further amplification in this configuration was restricted by the clear aperture of the isolator ($12$~mm), which constrained the laser beam size through the amplifier, thus preventing more efficient utilization of the amplifier rod volume. 

A second iteration was tested by using an expander capable of 2$\times$ magnification~\footnote{161-2X-1H, Eksma Optics}, positioned after the isolator, as shown in Fig.~\ref{fig:conf}(b). In this configuration, the collimated beam entering the amplifier was magnified to a larger diameter ($14$~mm - $1.9\times$ magnification).
This resulted in increased amplification evidenced by the maximum measured output pulse energy being enhanced to $1.8$~J for the same initial input energy (five-fold amplification).
However, the positioning of the beam expander after the isolator may lead to unwanted complications that could potentially damage the laser system.
It should be important to note that in this particular configuration, the reflections from beams along with the emissions from the amplifier can back-focus into the Faraday isolator, which could lead to damage over time.
To accommodate for this issue, the configuration described in section~\ref{sec:design} was incorporated.

\begin{backmatter}
\bmsection{Funding}
AG, SK, YZ, AK and GF are supported by
the Luxembourg National Research Fund 15480342 (FRAGOLA).
AG \& MK are supported by the Luxembourg National Research Fund 17838565 (ULTRAION).
AG \& AV are supported by the Luxembourg National Research Fund 17382436 (VERITAS).

\bmsection{Acknowledgment}
AG, SK, YZ, AK and GF are supported by
the Luxembourg National Research Fund 15480342 (FRAGOLA).
AG \& MK are supported by the Luxembourg National Research Fund 17838565 (ULTRAION).
AG \& AV are supported by the Luxembourg National Research Fund 17382436 (VERITAS).

\bmsection{Disclosures}
The authors declare no conflicts of interest.

\bmsection{Data Availability Statement}
The presented datasets are available upon request to the authors.


\end{backmatter}


\bibliography{5th_stage_laser}

\end{document}